\documentclass[graybox]{svmult}

\usepackage{mathptmx}       
\usepackage{helvet}         
\usepackage{courier}        
\usepackage{type1cm}        
\usepackage{makeidx}         
\usepackage{graphicx}        
\usepackage{multicol}        
\usepackage[bottom]{footmisc}

\makeindex             

\begin{document}

\title*{Low energy scales of Kondo lattices: mean-field perspective}
\author{S. Burdin}
\institute{S. Burdin \at Institut f\"ur Theoretische Physik, 
Universit\"at zu K\"oln, 
Z\"ulpicher Str. 77, 50937 K\"oln, Germany, 
\at Max Planck Institut f\"ur Physik Komplexer Systeme, 
N\"othnitzer Str. 38, 01187 Dresden, Germany, 
\email{burdin@thp.uni-koeln.de}}

%
%
\maketitle

\abstract{
A review of the low temperature properties of Kondo lattice systems 
is presented within the mean-field approximation, focusing on the 
different characteristic energy scales. 
The Kondo temperature, $T_{K}$, and the Fermi liquid coherence energy,
$T_{0}$, are analyzed as functions of the electronic filling, the 
shape of the non-interacting density of states, and the concentration 
of magnetic moments. 
These two scales can vanish, corresponding to a breakdown of the Kondo
effect when an external magnetic field is applied. The Kondo breakdown 
can also be reached by adding a superexchange term to the Kondo
lattice model, which mimics the intersite magnetic correlations
neglected at the mean-field level. 
}
\section{Introduction}
\label{sec:1}

Rare-earth and actinide based compounds exhibit extremely rich phase diagrams, 
with signatures of heavy fermion behavior, unconventional magnetism, or 
superconductivity~\cite{Fulde-Review, Matthias-Achim-Review}. 
At high temperature, the main physical properties of these systems 
are well reproduced by single impurity models, which describe the 
coupling between conduction electrons and one $4f$ or $5f$ ion. 
For dense systems, the single impurity models fail to describe the
low temperature properties, which are characterized by the formation 
of a non local coherent state. In this regime, 
models with a periodic lattice of $f$ ions are more appropriate. 

Here, we consider more specifically dense compounds where the $f$ orbital is 
occupied by one electron (Cerium) or one hole (Ytterbium).  
In the low temperature regime where the crystal field splitting lifts the
degeneracy of the $f$ orbital, these impurities are modeled by
effective local spins $S_{i}=1/2$. The system is thus described by the Kondo
lattice Hamiltonian, 
\begin{eqnarray}
H=\sum_{{\bf k}\sigma}(\epsilon_{_{\bf k}}-\mu)
c_{{\bf k}\sigma}^{\dagger}c_{{\bf k}\sigma}
+J_{K}\sum_{i}{\bf s}_{i}{\bf S}_{i}~, 
\label{KLHamiltonian}
\end{eqnarray}
where $c_{{\bf k}\sigma}^{\dagger}$ ($c_{{\bf k}\sigma}$) describe
creation (annihilation) operators of conduction electrons with spin 
$\sigma=\uparrow,\downarrow$ and momentum ${\bf k}$. 
The Kondo interaction results from a local antiferromagnetic coupling 
$J_{K}$ between the density of spin of conduction
electrons at site $i$, ${\bf s}_{i}$, and the Kondo impurities, ${\bf S}_{i}$. 
The chemical potential $\mu$ fixes the electronic filling to $n_{c}$ 
conduction electrons per site. 

\begin{figure}[h]
\includegraphics[width=7cm, origin=br, angle=90]{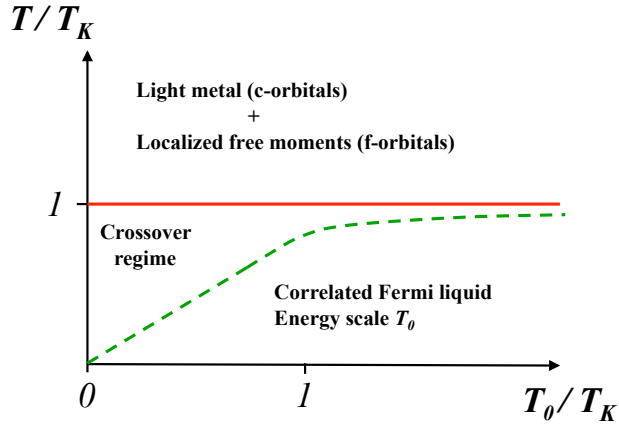}
\caption{Schematic phase diagram. The reduced temperature $T/T_{K}$ 
is plotted versus $T_{0}/T_{K}$, which is considered a tunable
parameter that can vary with the electronic filling, impurity
concentration, magnetic field, or the shape of the non-interacting
DOS. Figure from Ref.~\cite{BZ-2008}}
\label{Schemasgeneralphasediag}       
\end{figure}

\begin{figure}[h]
\includegraphics[width=4.5cm, origin=br, angle=-90]{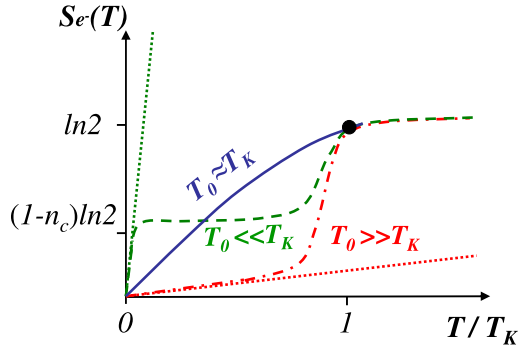}
\vspace*{0.7cm}
\caption{(Color online) Schematic plot of the electronic contribution to  
the entropy $S_{e^{-}}$ as a function of the normalized temperature  
$T/T_{K}$ for three cases: $T_{0}>>T_{K}$ (red dash dotted line), 
$T_{0}\approx T_{K}$ (blue solid line), and $T_{0}<<T_{K}$ 
(green dashed line). The dotted lines indicate  
the linear Fermi liquid regime $S_{e^{-}}(T)=T/T_{0}$, with a  
rescaled slope $T_{K}/T_{0}$.  
For $T>T_{K}$ the three curves are identical, 
reflecting the linear contribution from the conduction band 
$S_{e^{-}}(T)=\ln{2}+T/D$. 
The black dot refers to a standard  
experimental determination of $T_{K}$ from the electronic entropy:  
$S_{e^{-}}(T_{K})=p\ln{2}$. On this schematic plot, we used $p=1$,
which coincides with the vanishing of the effective hybridization,
$r=0$, defining $T_{K}$ within the 
mean-field approach. Experimentally, where $T_{K}$ is a crossover, 
one can choose, e.g., $p=1/2$. Figure from Ref.~\cite{BZ-2008}. } 
\label{Schemasentropy} 
\end{figure}

The Kondo model has been extensively studied throughout the last 
decades~\cite{Hewson}. At high temperature, the Kondo interaction can
be considered a small perturbation: conduction electrons and 
Kondo ions are weakly coupled. The transport properties, which are 
determined by the conduction band, correspond to those of a normal metal. 
The magnetic susceptibility, governed by the Kondo free moments, has a 
Curie-Weiss form. The entropy is large, of the order of $\ln 2$ per
site. 
A crossover occurs at the Kondo temperature, $T_{K}$, below which 
the Kondo interaction cannot be treated by perturbative methods. 
Experimental signatures of this crossover include, for example, a
logarithmic increase of the resistivity when the temperature
decreases. Other signatures involve a saturation of the magnetic
susceptibility, and a significant decrease of the entropy. 
At lower temperature, if we neglect magnetic or superconducting
instabilities, the physical properties are characteristic of a 
universal heavy Fermi liquid: the specific heat vanishes linearly 
with the temperature, 
$C_{V}(T)\approx \gamma T$; the local magnetic susceptibility, 
$\chi (T)$, as well as the resistivity are constant at $T=0$, with quadratic 
(i.e., $T^2$ like) variations at low $T$. 
This low temperature regime is characterized 
by an energy scale, the coherence temperature, $T_{0}$, which 
can be equivalently determined from the specific heat Sommerfeld 
coefficient, $T_{0}\equiv 1/\gamma$, or the zero temperature magnetic 
susceptibility, $T_{0}\equiv 1/\chi(T=0)$. 
Figure~\ref{Schemasgeneralphasediag} depicts the schematic phase
diagram of the model, as a function of $T_{0}/T_{K}$. 
The determination of $T_{0}$ and $T_{K}$ from the entropy is 
illustrated by Fig.~\ref{Schemasentropy}. 
The connection between $T_{0}$ and the thermal and electric transport 
properties have been analyzed by Zlati\`c 
{\it et al.}~\cite{Veljko1, Veljko2}. 

Note that the authors of Ref.~\cite{Pines} discussed two energy
scales: one of them corresponds to our definition of the Kondo 
lattice temperature, $T_{K}$, from the temperature dependence of the 
entropy [see Fig.~\ref{Schemasentropy}]. 
The second is the single impurity Kondo 
temperature, which, within the mean-field approach, is equal to the 
lattice Kondo temperature. 
The coherence temperature $T_{0}$ that we analyze here is not
considered in Ref.~\cite{Pines}. 

The physical properties of systems 
with a small concentration 
of magnetic ions are universal and characterized by a single energy scale,
$T_{0}=T_{K}$. The identity relating the coherence and the Kondo 
temperatures is consistent with the 
exact solution of the single impurity model. The situation is different 
for dense systems for which more than one energy scale can be identified. 
As an example, the Ruderman-Kittel-Kasuya-Yosida (RKKY) interaction, 
$J_{RKKY}$, can compete with $T_{K}$, and a magnetic instability can
be obtained for a lattice system~\cite{Doniach}. 
For a long time, it has been believed that $T_{0}$ and $T_{K}$ 
remain equal to each other 
in the heavy Fermi liquid phase of a Kondo lattice 
model. 
In this context and in response to discrepancies between theory and 
photoemission spectroscopy on rare-earth-based compounds 
with respect to the scaling with $T_{K}$, 
alternative scenarios involving phonons have been 
proposed~\cite{Photoemission1}. On this basis, whether 
{\it the Kondo model} is capable of predicting photoemission spectroscopy 
on YbAl$_{3}$ has been a subject of 
controversy~\cite{Photoemission2, Photoemission3, Photoemission4}. 
The specificity of Kondo lattice systems, with $T_{0}$ 
different 
from $T_{K}$, has been confirmed by experiments including magnetic 
susceptibility, specific heat, Hall coefficient measurements 
and X-ray absorption spectroscopy 
in Cerium, 
Ce$_{1-x}$La$_{x}$Ir$_{2}$Ge$_{2}$, 
CeIr$_{2-x}$(Rh,Pt)$_{x}$Ge$_{2}$, 
CeIr$_{2}$Ge$_{2-x}$(Si,Sn)$_{x}$~\cite{ExpeKondoalloy2}, 
CeNiSi$_{2}$~\cite{ExpeKondoalloy3}, and Ytterbium compounds, 
YbXCu$_{4}$œôòâ(X=Ag, Cd, In, Mg, Tl, Zn)~\cite{ExpeKondoalloy5}, 
Yb$_{1-x}$Lu$_{x}$Al$_{3}$~\cite{ExpeKondoalloy1}. 

The first suggestion that $T_{0}$ could be much smaller than 
$T_{K}$ was discussed by Nozi\`eres in the framework of the exhaustion 
problem~\cite{Nozieres-1985, Nozieres-1998}. This prediction
stimulated complementary theoretical works, including 
approximated methods based on a strong coupling
approach~\cite{Lacroix-1985} or 
mean-field calculations~\cite{
Lacroix-1986, BGG-2000, exhaustionandRKKY1, exhaustionandRKKY2}, and 
numerical simulations~\cite{
DMFT-exhaust1, DMFT-exhaust2, DMFT-exhaust3, DMFT-exhaust4,
DMFT-exhaust5} using dynamical mean-field 
theory (DMFT)~\cite{DMFT1, DMFT2}. Even if the initial prediction of
Nozi\`eres turned to be quantitatively 
wrong~\cite{BGG-2000, Nozieres-2005}, all the theoretical calculations
converge to the same qualitative conclusion: in a Kondo lattice, $T_{0}$ can be
different from $T_{K}$, and the ratio between these two energy scales
can be tuned, for example, by varying the electronic filling of the
system. 
Finally, these theoretical works shed some new light on the formerly 
controversial photoemission analysis. 

The aim of this work is, first, to review how the two energy scales, 
$T_{0}$ and $T_{K}$ depend on physical parameters: 
electronic filling, shape of the density of states, concentration of 
magnetic ions. This is done within the simplest relevant approximation
for the Kondo interaction: a mean-field decoupling. 
Then, we review how, within the mean-field approach, the Kondo phase 
can be destabilized by a magnetic field or by magnetic inter-ion 
interactions.

\section{Mean-field formalism}
\label{sectionmeanfield}
The mean-field approximation for the Kondo lattice was first 
introduced by Lacroix and Cyrot~\cite{Lacroix-Cyrot}. 
It was reformulated by Coleman~\cite{Coleman-1983} 
and by Read, Newns and Doniach~\cite{Read-Newns-Doniach-1984}, 
as a large$-N$ approximation for 
the $N-$fold degenerate Coqblin-Schrieffer 
model~\cite{Coqblin-Schrieffer}. 
It was shown that magnetic instabilities require expansions 
up to the order $1/N$, i.e. fluctuations around the mean-field. 
Nevertheless, the heavy Fermi liquid phase is well described in the 
limit $N=\infty$. The analysis of $T_{0}$ and $T_{K}$ can thus 
be already performed at the mean-field level. Here, we describe 
the main lines of the mean-field approximation for the 
Kondo lattice Hamiltonian~(\ref{KLHamiltonian}). 

The Kondo impurities are represented by local 
auxiliary fermions as follows: 
$S_{i}^{z}=\frac{1}{2}(f_{i\uparrow}^{\dagger}f_{i\uparrow}
-f_{i\downarrow}^{\dagger}f_{i\downarrow})$, 
$S_{i}^{+}=f_{i\uparrow}^{\dagger}f_{i\downarrow}$, and 
$S_{i}^{-}=f_{i\downarrow}^{\dagger}f_{i\uparrow}$. 
The Kondo interaction is thus rewritten as 
$J_{K}{\bf s}_{i}{\bf S}_{i}\mapsto 
\frac{J_{K}}{2}\sum_{\sigma\sigma'}c_{i\sigma}^{\dagger}c_{i\sigma'}
f_{i\sigma'}^{\dagger}f_{i\sigma}$, which describes the spin-flip
processes between conduction electrons and local moments. 
This mapping is exact as far as the Hilbert space
is restricted to the sector of one auxiliary fermion per site, 
$f_{i\uparrow}^{\dagger}f_{i\uparrow}
+f_{i\downarrow}^{\dagger}f_{i\downarrow}=1$. 
The mean-field solution is obtained within 
the two following approximations: 
{\it (i)} The local occupation of the
auxiliary fermions is equal to one only on average. 
This corresponds to describing 
the Kondo spins by an effective local $f-$level that is half full. 
This effective filling is controlled by introducing 
a second chemical potential, $\lambda$. 
{\it (ii)} 
The Kondo interaction is replaced by an effective one-body 
term, obtained from a mean-field decoupling of the 
$\uparrow$ and $\downarrow$ components. 
The mean-field approximation for the Kondo lattice
Hamiltonian~(\ref{KLHamiltonian}) yields  
\begin{eqnarray}
H=\sum_{{\bf k}\sigma}(\epsilon_{_{\bf k}}-\mu)
c_{{\bf k}\sigma}^{\dagger}c_{{\bf k}\sigma}
+
r\sum_{i\sigma}[c_{i\sigma}^{\dagger}f_{i\sigma}+f_{i\sigma}^{\dagger}c_{i\sigma}]
-\lambda\sum_{i\sigma}f_{i\sigma}^{\dagger}f_{i\sigma}~, 
\label{MFHamiltonian}
\end{eqnarray}
where the effective hybridization is determined by the self-consistent 
relation 
\begin{eqnarray}
r=\frac{J_{K}}{2{\cal N}}\sum_{i\sigma}
\langle f_{i\sigma}^{\dagger}c_{i\sigma}\rangle~.  
\end{eqnarray}
Here ${\cal N}$ is the number of lattice sites. 
The chemical potentials for $c-$electrons, $\mu$, and $f-$fermions, 
$\lambda$, are determined by the constraints 
$n_{c}=\frac{1}{\cal N}\sum_{i\sigma}
\langle c_{i\sigma}^{\dagger}c_{i\sigma}\rangle$ and 
$1=\frac{1}{\cal N}\sum_{i\sigma}
\langle f_{i\sigma}^{\dagger}f_{i\sigma}\rangle$. 
The mean-field Hamiltonian~(\ref{MFHamiltonian}) describes an
effective system of conduction electrons hybridized with local $f$
levels. The correlation effects are renormalized into the self-consistent 
hybridization $r$. 
This approximation captures two important features of the Kondo lattice: at high
temperature, we find $r=0$, and the system is described as a
paramagnetic light metal ($c$ electrons) decoupled from local free
moments ($f-$fermions). This picture is oversimplified but it succeeds 
in describing qualitatively  
the experimental situation, in which, above the Kondo 
temperature, conduction electrons are weakly coupled to the local
moments. Within the mean-field approach, the Kondo temperature $T_{K}$ is thus
defined as the temperature for which a hybridization $r\neq 0$
occurs. 
At very low temperature, the physical properties correspond to a Fermi
liquid and the excitations of the system correspond to 
the creation of non-interacting, heavy, fermionic quasiparticles. The latter
are a linear combination of the light $c-$electrons and the heavy 
$f-$fermions. 
The Fermi-liquid regime is characterized by an energy scale,
$T_{0}$, which can be defined identically from different physical
properties of the ground state: 
the quasiparticle density of states, $\rho=1/T_{0}$, 
the Sommerfeld coefficient, $\Gamma=lim_{T\to 0}C_{V}(T)/T=1/T_{0}$, 
or the local impurity spin susceptibility, $\chi_{loc}(T=0)=1/T_{0}$. 
We have obtained explicit expressions for $T_{K}$ and $T_{0}$ in the 
limit of small $J_{K}$~\cite{BZ-2008, BGG-2000}: 
\begin{eqnarray} 
\label{ExpressionTK}
T_{K}=F_{K}[n_{c},\rho_{0}]e^{-1/J_{K}\rho_{0}(\mu_{0})}~, 
\end{eqnarray}
and 
\begin{eqnarray}
\label{ExpressionT0}
T_{0}=F_{0}[n_{c},\rho_{0}]e^{-1/J_{K}\rho_{0}(\mu_{0})}~, 
\end{eqnarray}
which depend on the Kondo coupling $J_{K}$ only within the 
non-analytic exponential factor 
$e^{-1/J_{K}\rho_{0}(\mu_{0})}$. 
The prefactors $F_{K}$ and $F_{0}$
are functions of the electronic filling, $n_{c}$, and the
non-interacting density of states (DOS),  
$\rho_{0}(\omega)\equiv\sum_{\bf k}\delta(\omega-\epsilon_{\bf k})$. 
Here, $\mu_{0}$ is the 
non-interacting chemical potential, corresponding to an electronic
filling 
$n_{c}=\int_{-D}^{\mu_{0}}\rho_{0}(\omega)d\omega$. 
For the sake of clarity, we do not write the explicit 
expressions of $F_{K}$ and $F_{0}$ here, as they are given 
in Refs.~\cite{BZ-2008, BGG-2000}. 
The exponential factor in both $T_{0}$ and $T_{K}$, 
results in a very sensitive dependence of these energy scales with
respect to small changes in the system (e.g., pressure or doping). 
Since the ratio $T_{0}/T_{K}$ does not depend on
$J_{K}$, it is more likely to be analyzed experimentally as a
universal function. 
The mean-field solution 
yields~\cite{BZ-2008} 
\begin{eqnarray}
\frac{T_{0}}{T_{K}}=
\left( \frac{D+\mu_{0}}{D-\mu_{0}}\right)^{1/2}
\frac{F_{shape}}{\alpha \rho_{0}(\mu_{L})\Delta\mu}~. 
\label{TzerosurTK}
\end{eqnarray}
Here, $D$ is the half-bandwidth of the non-interacting DOS. 
The interacting
chemical potential, $\mu_{L}$, corresponds to a filling $n_{c}+1$, and the energy 
$\Delta\mu\equiv\mu_{L}-\mu_{0}$ is related to the 
enlargement of the Fermi surface. $\alpha=1.13$ is a number. 
$F_{shape}$ is an explicit function depending on the shape of
$\rho_{0}$ as follows: 
\begin{eqnarray}
\label{Fshape}
F_{shape}\equiv\exp{\left[ 
\left(
\int_{-(D+\mu_{0})}^{\Delta\mu}
-\frac{1}{2}\int_{-(D+\mu_{0})}^{D-\mu_{0}}
\right)
\frac{\rho_{0}(\mu_{0}+\omega)-\rho_{0}(\mu_{0})}{\vert\omega\vert 
\rho_{0}(\mu_{0})}d\omega
\right]}~. 
\end{eqnarray}

\section{Tuning $T_{0}$ and $T_{K}$}

\subsection{Variation of electronic filling}
\label{Section-filling}
The effect of electronic filling on the temperature scales of the
Kondo lattice has been discussed by 
Nozi\`eres who suggested a possible exhaustion 
problem~\cite{Nozieres-1985, Nozieres-1998}. 
Considering $N_{S}$ Kondo spins coupled to 
$N_{c}\le N_{S}$ conduction electrons, 
Nozi\`eres started with the following remark: the magnetic 
entropy which can be released at the temperature $T\approx T_{K}$ by the formation of
incoherent Kondo singlets (i.e., by the single impurity Kondo effect) 
is $\Delta S = N_{c}\ln 2$. 
However, the formation of a coherent Fermi liquid ground state is 
characterized by a vanishing entropy, $S(T)\approx T/T_{0}$. 
The freezing of the remaining entropy, 
$S(T_{K})\approx (N_{S}-N_{c})\ln2$, 
thus requires a collective mechanism and can lead to 
$T_{0}\ll T_{K}$ when $N_{c}\ll N_{S}$. 

\begin{figure}[hhh]
\includegraphics[width=6cm, origin=br,angle=-90]{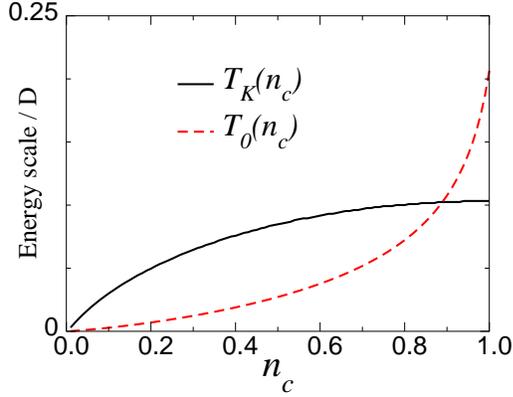}
\caption{(Color online) 
$T_{K}/D$ (black solid line) and $T_{0}/D$ (red dashed line) versus electronic filling, 
for the Kondo lattice. 
Numerical result obtained within the mean-field approximation, 
for a semi-elliptic non-interacting DOS and 
$J_{K}/D=0.75$~\cite{BGG-2000}. }
\label{TzeroetTKdenc}
\end{figure}

The exhaustion problem remained for several years an open issue, and
its first solution was obtained within a mean-field 
approach~\cite{BGG-2000, Nozieres-2005}, which 
provides a quantitative description of the electronic filling effect. 
First, from the analytical 
expressions~(\ref{ExpressionTK}-\ref{ExpressionT0}), $T_{K}$ and
$T_{0}$ have the same $J_{K}$ dependence,
$e^{-1/J_{K}\rho_{0}(\mu_{0})}$ 
factor, and 
the ratio $T_{K}/T_{0}$ does not depend on the Kondo 
coupling. 
This is in contradiction with the result of 
Nozi\`eres, who predicted a ratio 
$T_{0}/T_{K}\approx T_{K}/D\approx 
e^{-1/J_{K}\rho_{0}(\mu_{0})}$~\cite{Nozieres-1985, Nozieres-1998}. 
The mean-field result, $T_{0}/T_{K}$ independent of $J_K$, 
was later confirmed by 
DMFT calculations~\cite{Costi-Manini}, and 
finally accepted by Nozi\`eres~\cite{Nozieres-2005}. 
Nevertheless, $T_{0}$ and $T_{K}$ define two energy scales 
with different dependencies with respect to 
the electronic filling, as depicted by 
Fig.~\ref{TzeroetTKdenc}, where $T_{0}\ll T_{K}$ in the 
limit $n_{c}\to 0$. The mean-field result obtained here 
is remarkably similar to the one obtained within DMFT combined 
with Quantum Monte Carlo simulation [see Fig.~1 in Ref.~\cite{DMFT-exhaust3}].  
The filling effects can also be analyzed from the 
expression~(\ref{TzerosurTK}). 
Neglecting the band shape effects 
[discussed in section~\ref{sectionshape}], 
we find $T_{0}/T_{K}\approx \left(
  \frac{D+\mu_{0}}{D-\mu_{0}}\right)^{1/2}$, 
which vanishes when $\mu_{0}$ approaches the band edge $-D$, i.e.,
when $n_{c}\to 0$. 

\begin{figure}[hhh]
\includegraphics[width=6cm, origin=br,angle=-90]{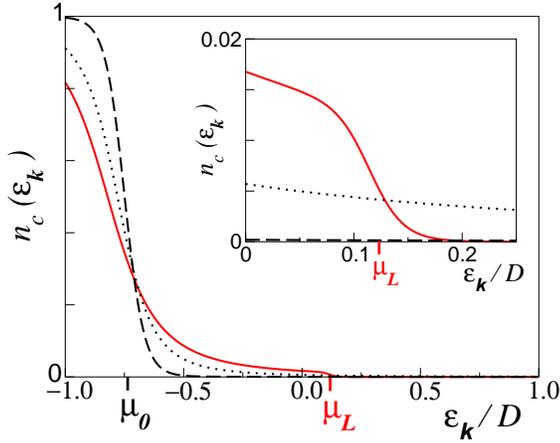}
\caption{(Color online)
Electronic occupation $n_{c}(\epsilon_{\bf k})$ for $T/T_{K}=1.0$ 
(black dashed line), $T/T_{K}=0.5$ (black dotted line), and
$T/T_{K}=0.005$ (red solid line). 
Numerical result obtained within the 
mean-field approximation for a semi-elliptic non-interacting DOS, 
$n_{c}=0.15$, and $J_{K}/D=0.75$. $\mu_{0}$ and $\mu_{L}$ indicate 
the chemical potential corresponding to a small and a large Fermi surface, 
respectively. Inset: focus around $\mu_{L}$. 
Figure from Ref.~\cite{BGG-2000}. 
 }
\label{occupation}
\end{figure}

The analysis of the electronic occupation, 
$n_{c}(\epsilon_{\bf k})\equiv \langle c_{\bf k}^{\dagger}c_{\bf k}\rangle$, 
provides an important insight for understanding the physical mechanism 
leading to two different energy scales. 
For $T\approx T_{K}$, the mean-field result, 
depicted by Fig.~\ref{occupation}, looks like a Fermi 
distribution with a thermal window 
around the non-interacting chemical potential, $\mu_{0}$, 
corresponding to $n_{c}$ electrons (small Fermi surface). 
For $T\ll T_{K}$, in the Fermi liquid regime, the distribution is
spread and forms a step around $\mu_{L}$ (large Fermi surface). 
Figure~\ref{occupation} only describes the occupation of the
$c-$electrons, which is fixed to $n_{c}$. 
At $T=0$, there are states of given momentum ${\bf k}$ 
which are not fully occupied by  
$c-$electrons. The Fermi liquid picture is recovered because the
quasiparticles are not pure $c-$states, but a linear combination of 
$c$ and $f-$states. The quasiparticle occupation is complete, i.e.,
equal to one, for states with an energy $\epsilon_{\bf k}<\mu_{L}$,
and it vanishes for states with higher energy. 
This behavior is consistent with the Luttinger theorem which predicts
that, at $T=0$, the Fermi surface contains 
both $c-$electrons and $f-$fermions. It is not surprising that the
Kondo lattice satisfies the Luttinger theorem within the mean-field 
approximation. The reason is that the effective mean-field 
model~(\ref{MFHamiltonian}) does not contain an explicit many body
interaction term. We expect this result to survive beyond the mean-field 
in the Fermi liquid phase. 
The enlargement of the Fermi surface might be a key point in the
origin of the difference between $T_{0}$ and $T_{K}$: the Kondo
temperature is associated with the incoherent scattering of 
the conduction electrons which are in the Fermi window of width
$T_{K}$ around $\mu_{0}$ [see Fig.~\ref{Schemasdos}]. The coherence 
temperature, $T_{0}$, results from the Kondo effect, but, unlike
$T_{K}$, characterizes a Fermi liquid with a large Fermi surface.

\subsection{Shape of the non-interacting density of states}
\label{sectionshape}

\begin{figure}[h]
\includegraphics[width=2.5cm, origin=br, angle=-90]{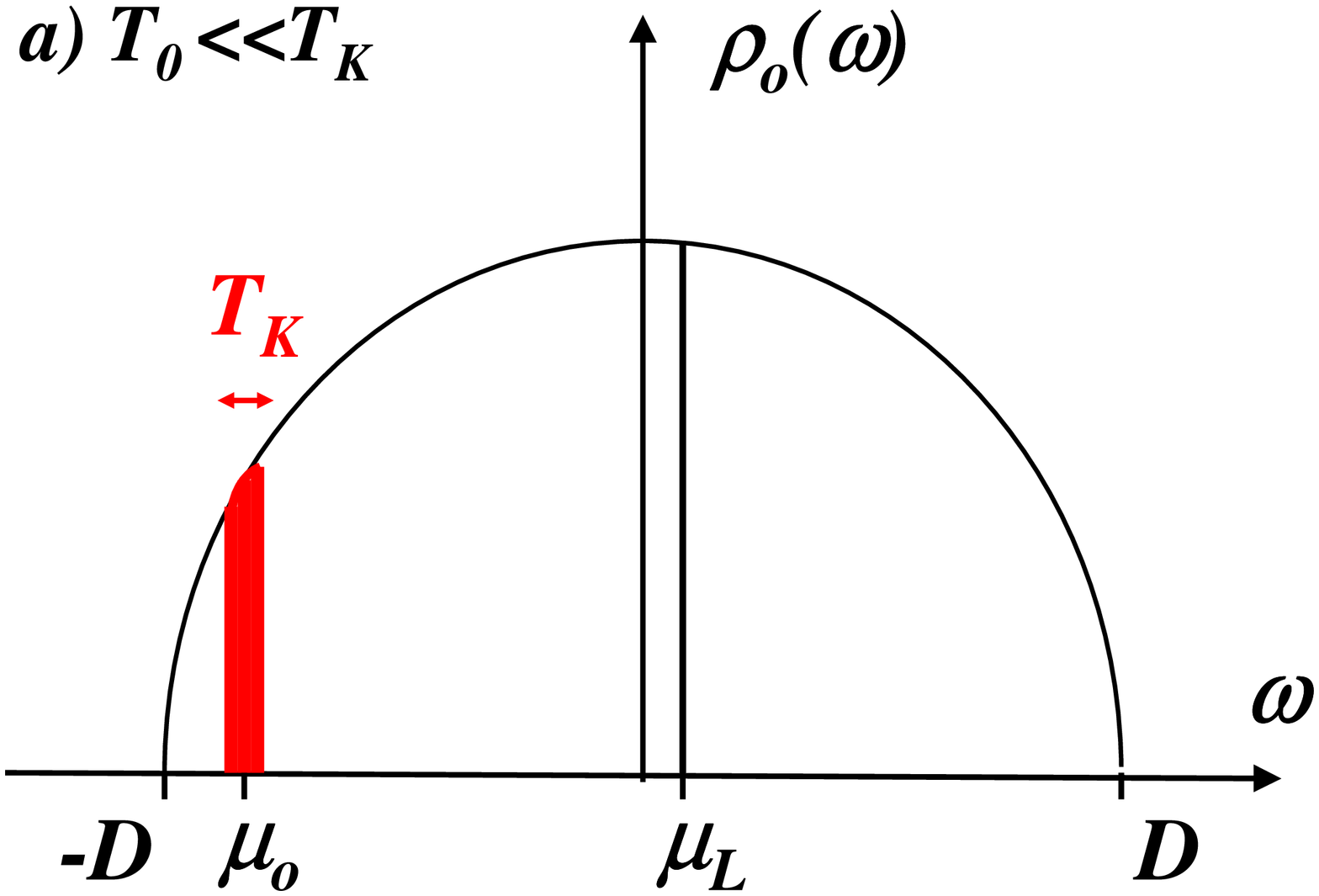}~~~
\includegraphics[width=2.5cm, origin=br, angle=-90]{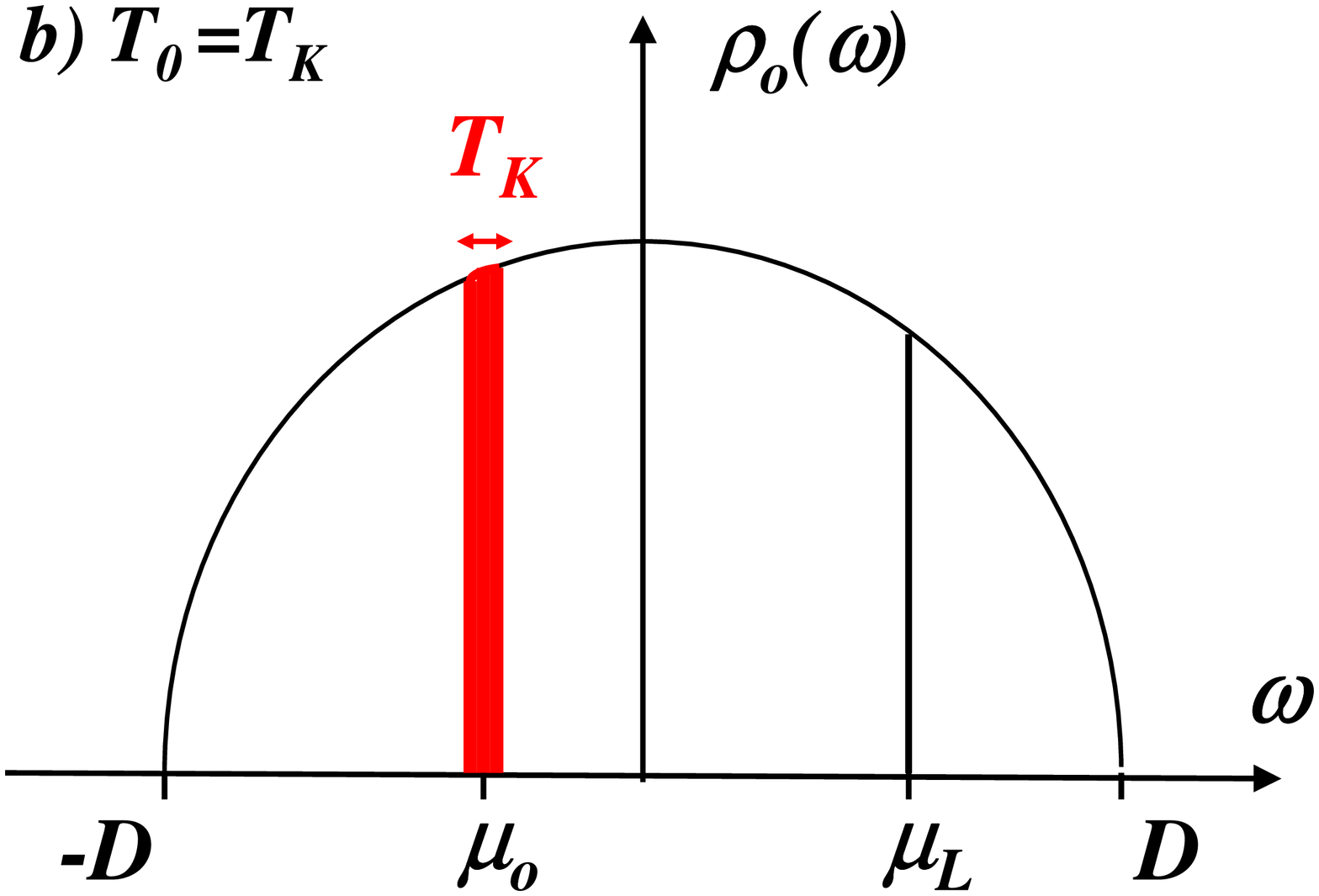}~~~
\includegraphics[width=2.5cm, origin=br, angle=-90]{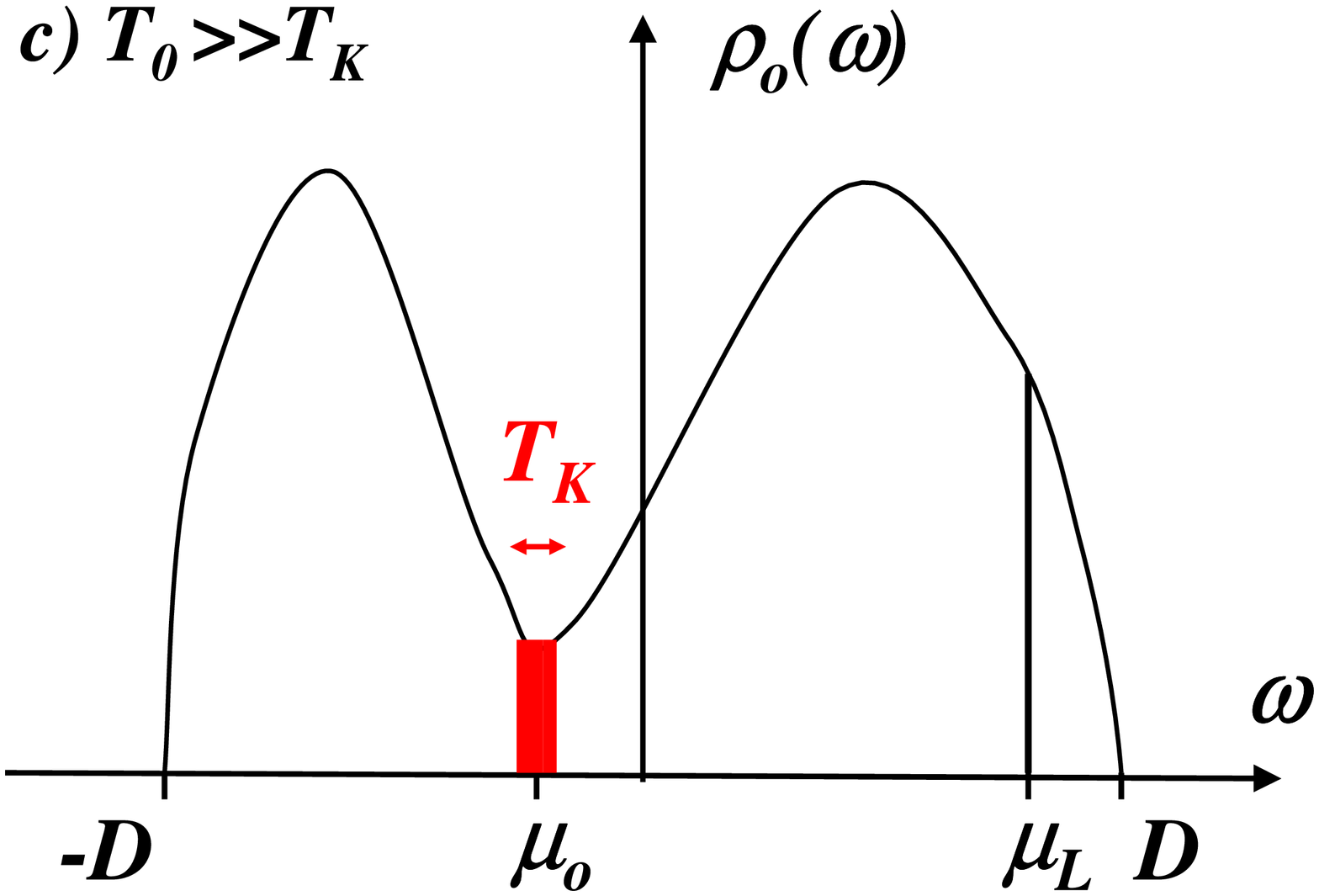}
\caption{
(Color online) Schematic plot of the non-interacting DOS. (a) For a 
regular DOS and far from the electronic half-filling $T_{0}\ll T_{K}$. 
Close to the half-filling, the shape of the DOS around $\mu$ is 
crucial for determining $T_{0}/T_{K}$: (b) $\rho_{0}(\omega)$ is 
nearly constant around $\omega=\mu$ and $T_{0}\sim T_{K}$. 
(c) $\mu$ is close to a minimum of $\rho_{0}$ and $T_{0}\gg T_{K}$. 
Here, $\mu_{0}\approx \mu$ indicates the chemical potential
corresponding to $n_{c}$ non-interacting $c-$electrons (small
Fermi surface), and $\mu_{L}$ is the chemical potential corresponding
to $n_{c}+1$ non-interacting $c-$electrons (large Fermi surface). 
Figure from Ref.~\cite{BZ-2008}. 
}
\label{Schemasdos}       
\end{figure}
Here we consider the effects due to the shape variation of the 
non-interacting DOS, $\rho_{0}(\omega)$~\cite{BZ-2008}. 
In order to separate this effect from the
electronic filling effects, we assume that $n_{c}$ is close to $1$,
but not exactly half full, so that the system is metallic. 
Eqs~(\ref{TzerosurTK}) and (\ref{Fshape}) are thus simplified to 
\begin{eqnarray}
\frac{T_{0}}{T_{K}}\approx F_{shape}\approx 
\exp\left[ \int_{-(D+\mu_{0})}^{D-\mu_{0}}
\frac{\rho_{0}(\mu_{0}+\omega)-\rho_{0}(\mu_{0})}{2\vert\omega\vert 
\rho_{0}(\mu_{0})}d\omega\right]~. 
\label{Ratioshape}
\end{eqnarray}
A constant $\rho_{0}$ gives $T_{0}\sim T_{K}$, which explains the
$T/T_{K}$ scaling observed in some compounds. 
If $\mu_{0}$ is close to a local maximum of $\rho_{0}(\omega)$, the
integrand in Eq.~(\ref{Ratioshape}) is negative in the main part of
the integration range, and $T_{0}\ll T_{K}$. 
In the opposite situation, when $\mu_{0}$ is close to a local minimum 
[see Fig.~\ref{Schemasdos}~(c)], 
we find $T_{0}\gg T_{K}$, which can be understood by the following
argument. 
The incoherent Kondo screening which begins at $T\approx T_{K}$ 
involves a small number of conduction electrons which are in the 
Fermi thermal window of width $T_{K}$, around $\mu_{0}$. 
At lower temperature, $T<T_{K}$, the Fermi surface is enlarged, due to 
the contribution of the $f-$electrons [see Fig.~\ref{occupation}]. 
This results from a non-zero 
hybridization, $r\neq 0$, in the Kondo phase. The formation of a 
coherent Fermi liquid ground state thus involves all the states of the 
large Fermi surface. In the situation described by 
Fig.~\ref{Schemasdos}~(c), where $\mu_{0}$ is close to a local minimum
of the DOS, the further we are from $\mu_{0}$, the more states are
available for the formation of the coherent Fermi liquid. 
In this case, there is a kind of self-amplification resulting in 
$T_{0}\gg T_{K}$. With this picture, the magnetic screening of the Kondo
impurities involves not only the $c-$conduction electrons but also, in
a dynamical way, "growing" quasiparticles which contain
some $f-$components.

\subsection{Substitution of magnetic ions}
Here, our analysis focuses on the concentration of Kondo 
impurities, $x$. In rare-earth based Kondo systems, 
the latter can be varied, e.g., by Ce-La or Yb-Lu substitution. 
For this purpose, we have introduced a Kondo alloy model, which is a 
generalization of the Hamiltonian~(\ref{KLHamiltonian}) 
where each site of the lattice can randomly either contain a Kondo 
impurity, with a probability $x$, or 
not~\cite{BF-2007, Matthias-2007}. 
The Kondo interaction is then treated within the mean-field 
approximation, and the different configurations of impurity 
distributions are averaged using a generalization of 
the coherent  potential approximation~\cite{MatrixCPA}. 
Results described here remain valid beyond the coherent 
potential approximation~\cite{Matthias-2007}. 

\begin{figure}[hhh]
\includegraphics[width=6cm, origin=br,angle=-90]{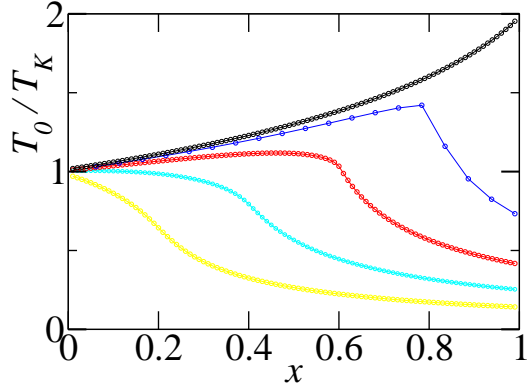}
\caption{(Color online) $T_{0}/T_{K}$ as a function of the impurity 
concentration. From top to bottom, $n_{c}=1;0.8;0.6;0.4;0.2$. The 
solid lines are included as a guide for the eye. 
The curves have been computed for
a semi-elliptic non-interacting DOS and $J_{K}/D=0.75$. 
From Ref.~\cite{BF-2007}. 
}
\label{TzerosurTKalloy}
\end{figure}

Figure~\ref{TzerosurTKalloy} illustrates the 
evolution of $T_{0}/T_{K}$ as a function of the impurity
concentration, $x$,  
for different values of electronic filling, $n_{c}$. The dilute limit of the
model, $x\ll 1$, reproduces the universal behavior of 
the single impurity model, with $T_{0}=T_{K}$. The crossover to the 
dense Kondo-lattice regime, with $T_{0}\neq T_{K}$ occurs for
$x\approx n_{c}$. 

\begin{figure}[hhh]
\includegraphics[width=6cm, origin=br,angle=-90]{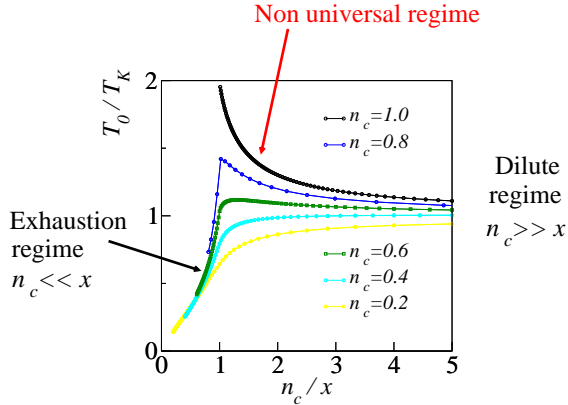}
\caption{(Color online) Same data as Figure~\ref{TzerosurTKalloy}, 
with a rescaled horizontal axis which represents $n_{c}/x$ here. 
Each curve corresponds to a given electronic filling, $n_{c}$, 
as indicated by
the legend. The parameter $n_{c}/x$ has been tuned by varying $x$, and
the curves are cut-off by the finite physical minimal value $n_{c}/x=n_{c}$. 
}
\label{TzerosurTKalloyrescaled}
\end{figure}

The data presented in Fig.~\ref{TzerosurTKalloyrescaled} are
identical to those in Fig.~\ref{TzerosurTKalloy}, with a rescaled 
axis, $x\to n_{c}/x$. 
From this rescaling, an exhaustion regime is identified, 
for $n_{c}<x$, where $T_{0}/T_{K}$ is a universal function of $n_{c}/x$. 
Note that this numerical result has been obtained here for a non-interacting
semi-elliptic DOS which mimics, at low $n_{c}$, the band edge of a 
three-dimensional system. We expect the universal exhaustion regime to 
depend on the dimension via the exponent characterizing the vanishing
of the DOS at the band edge. 
Apart from the possibility of magnetic ordering, which is not
considered here, the exhaustion regime might be difficult to access
experimentally for the following reason: the maximal concentration 
of Kondo impurities is $x=1$. The only way to reach $n_{c}\ll x$ thus
involves decreasing the electronic filling. Yet, in rare-earth
compounds, conduction electrons usually involve more than one band.  
With a single band description, the exhaustion regime would be
observable by increasing the concentration of magnetic ions, $x$. 
A multi-band system, however, allows for another scenario 
to materialize as $x$ increases: 
instead of exhausting one conduction band, the system might 
energetically prefer transferring electrons from other bands, preventing
the filling from accessing the exhaustion regime, $n_{c}\ll x$. 
In this case, 
a single band Kondo lattice model would not be appropriate anymore. 
Nevertheless, we expect that other Kondo systems can be
realized experimentally, 
in which the universal regime $n_{c}\ll x$ could be observed. 
For most of dense compounds, $x\approx 1$, the electronic filling 
is $n_{c}\approx x$. This is a non-universal regime, where all the
lattice structure becomes relevant.

\section{Destabilizing the Kondo phase}
Microscopically, the Kondo effect is characterized by the formation of
local singlets. The Kondo phase can be destabilized by an external
magnetic field, or by the fluctuations of 
the internal Weiss field induced by RKKY
interactions. Here, these two situations are analyzed from 
the mean-field approximation: the breakdown of the Kondo effect is  
identified to a continuous vanishing of the $f-c$ effective hybridization, $r$,
which occurs at $T=0$. 

\subsection{Effect of a magnetic field}
We consider an external magnetic field, $h$, applied to the Kondo spins
in the longitudinal direction $z$. 
This mimics a situation where the Land\'e
factor of the magnetic ions is much larger than the one of
$c-$electrons. 
A supplementary contribution is thus added to the 
Hamiltonian~(\ref{KLHamiltonian}), $H\mapsto H-h\sum_{i}S_{i}^{z}$. 
The mean-field approximation described in 
Sec.~\ref{sectionmeanfield} is generalized~\cite{BZ-2008}, 
resulting in 
an effective Hamiltonian formally similar to
Eq.~(\ref{MFHamiltonian}), with spin-dependent $f-$fermion 
potentials, $\lambda\mapsto \lambda+\sigma h$. 
The effective hybridization, $r$, remains spin-independent. 

\begin{figure}[hhh]
\includegraphics[width=3.5cm, origin=br,angle=-90]{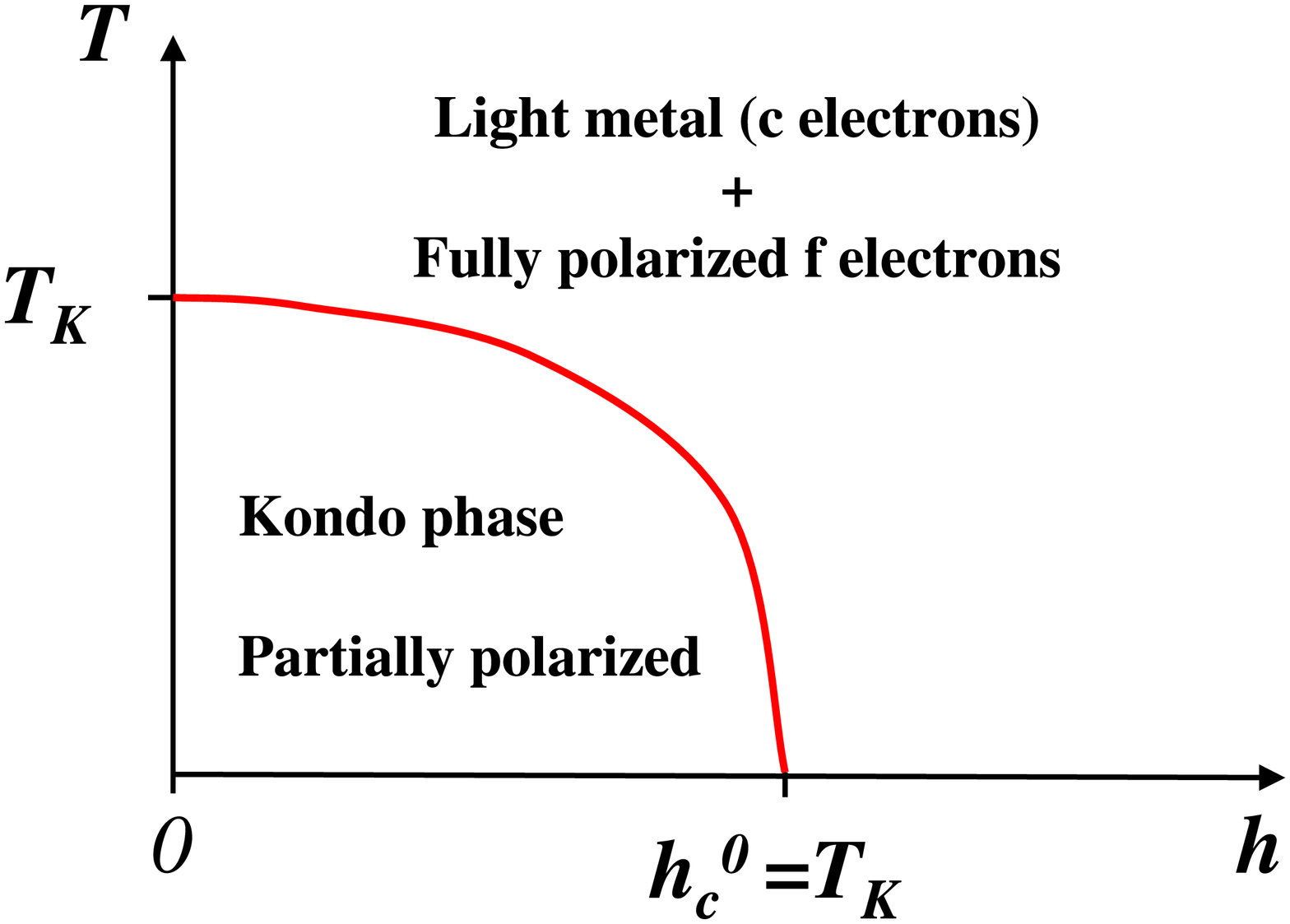}
\hspace*{1cm}
\includegraphics[width=3.5cm, origin=br,angle=-90]{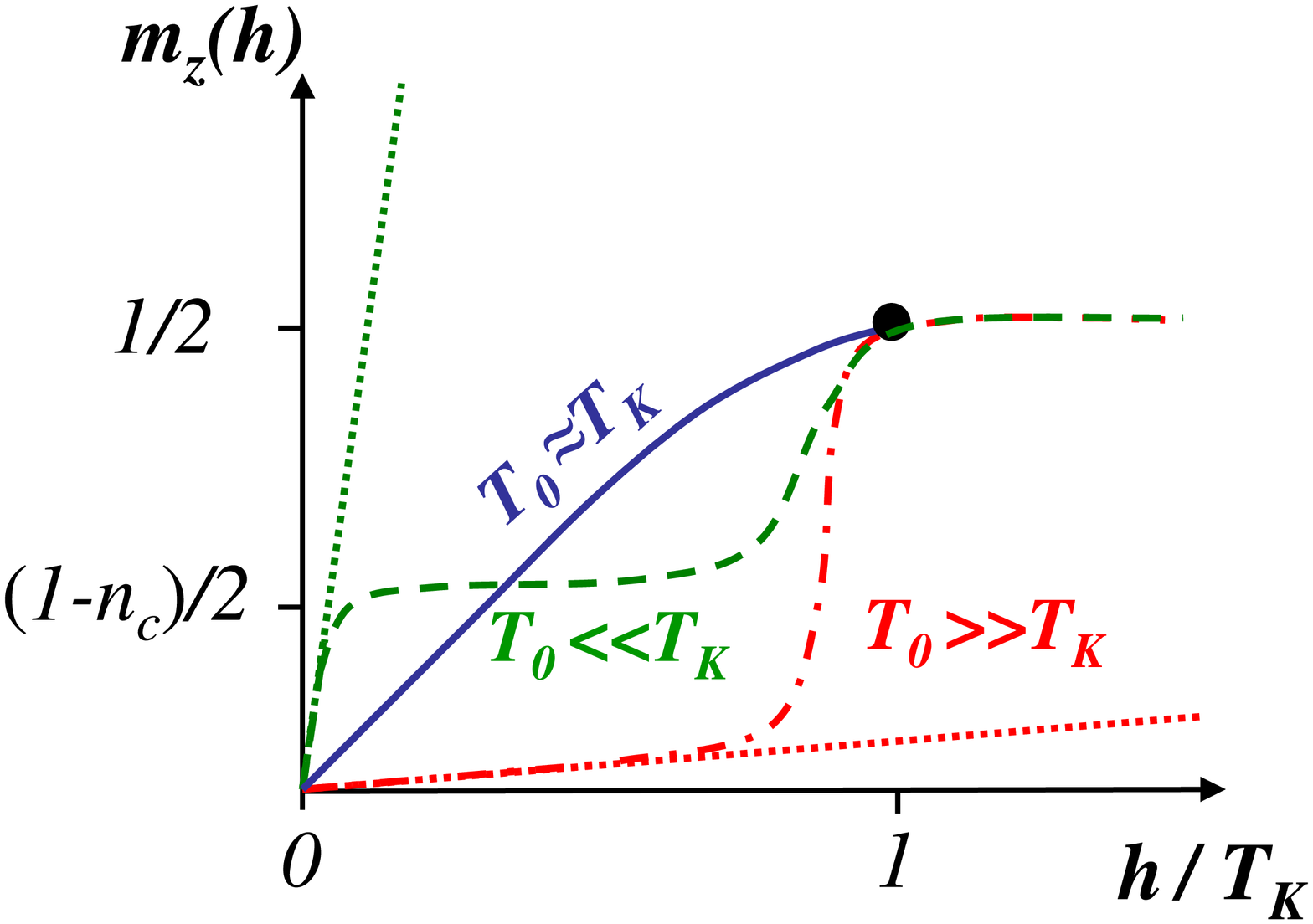}
\vspace*{0.7cm}
\caption{(Color online) 
Left: Schematic phase diagram of the Kondo lattice 
as a function of a magnetic field $h$. 
The red solid line indicates $h_{c}(T)$ which separates the Kondo 
phase ($r\neq 0$) from the 
decoupled phase ($r=0$). 
Right: Schematic plot of the magnetization 
$m_{z}(h/T_{K})$. For $T_{0}\gg T_{K}$ (red dash dotted line), 
$T_{0}\approx T_{K}$ (blue solid line), 
and $T_{0}\ll T_{K}$ (green dashed line). 
The dotted line indicates the initial slope in the linear response
regime, where $m_{z}(h)=h/T_{0}$. The black dot refers to the complete 
polarization of the local Kondo spins, $m_{z}=1/2$,  
which occurs at the critical field $h_{c}^{0}=T_{K}$. 
Figures from ref.~\cite{BZ-2008}. 
}
\label{Schemasmagnet}
\end{figure}

For a  sufficiently small magnetic field, a low temperature 
solution with $r\neq 0$ is obtained. This situation, depicted by
Fig.~\ref{Schemasmagnet}, characterizes a phase where the 
Kondo effect coexists with a partial polarization of both Kondo spins
and $c-$electrons. A finite critical field $h_{c}(T)$ is obtained,
above which $c-$electrons decouple from fully polarized local
moments. 
At $T=0$, in the weak coupling limit, $J_{K}\ll D$, the mean-field 
approximation yields the universal relation 
$h_{c}^{0}\equiv h_{c}(T=0)= T_{K}/\alpha$. 
Since $\alpha=1.13$, we have $h_{c}^{0}\approx T_{K}$. 
This result is not surprising if we consider the Kondo temperature as 
the energy scale characterizing the local singlet formation: 
the Kondo effect is destroyed when the Zeeman energy becomes larger than 
$T_{K}$. 
At finite temperature, assuming a constant $c-$DOS yields the
critical line $[h_{c}(T)/h_{c}^{0}]^{2}+[T/T_{K}]^{2}=1$. 


The $T=0$ magnetization $m_{z}(h)\equiv\frac{1}{2N}\sum_{i}
(\langle f_{i\uparrow}^{\dagger}f_{i\uparrow}\rangle
-\langle f_{i\downarrow}^{\dagger}f_{i\downarrow}\rangle )$ 
obtained from the mean-field solution is plotted in 
Fig.~\ref{Schemasmagnet} as a function of the reduced 
magnetic field $h/T_{K}$. 
The low field regime is given by the linear response, 
$m_{z}(h)=h\chi_{loc}(T=0)$, which, by definition, yields 
$m_{z}(h)=h/T_{0}$. Since the critical field characterizing a full 
polarization of the local moments is of the order of the Kondo 
temperature, $h_{c}^{0}\approx T_{K}$, we distinguish three typical 
cases. $T_{0}\approx T_{K}$ is a standard situation, where 
$m_{z}$ increases linearly with $h$, until saturation. 
For $T_{0}\gg T_{K}$, the linear regime allows only a small 
magnetization when $h< h_{c}^{0}$. Thus, at about 
$h\approx h_{c}^{0}$, such systems exhibit a meta-magnetic transition 
from an unpolarized Fermi-liquid to the polarized spin lattice. 
In the opposite case, $T_{0}\ll T_{K}$, the linear regime saturates 
around $h\sim T_{0}\ll h_{c}^{0}$. The intermediate regime
$T_{0}<h<T_{K}$ is expected to be non-universal. Eventually, a
magnetization plateau can occur, at $m_{z}\approx (1-n_{c})/2$, 
similar to the one obtained for the entropy 
[see Fig.~\ref{Schemasentropy}]. 

It is well known that the transition $r=0$, defining $T_{K}$
within the mean-field approximation, becomes a crossover 
when more accurate methods are used. 
There are several examples of systems where a finite temperature 
crossover ends up to a quantum critical point, i.e. 
a transition at zero temperature. 
Wether the transition at the critical field $h_{c}^{0}$ 
would survive beyond the mean-field is an open issue.

\subsection{Extra-RKKY interaction}
The possibility of destabilizing a Kondo phase in favor of a
magnetically ordered ground state was first discussed by 
Doniach~\cite{Doniach}, by comparison of the Kondo temperature, 
$T_{K}\sim De^{-1/J_{K}\rho_{0}(\mu_{0})}$, with the 
RKKY energy, 
$J_{RKKY}\sim \rho_{0}(\mu_{0})J_{K}^{2}$. 

The mean-field approximation for the Kondo lattice 
Hamiltonian~(\ref{KLHamiltonian}) can hardly provide a correct 
description of a transition from a Kondo ground state to a
phase with magnetic ordering. This results from the two following 
reasons: \\
{\it (i) Difficulty in driving the system to a second order
magnetic transition:} 
this would require a mechanism leading to a continuous 
vanishing of the zero temperature effective hybridization, $r(T=0)$. 
However, from the Kondo lattice 
Hamiltonian~(\ref{KLHamiltonian}), the only possibility of vanishing 
$r(T=0)$ involves the non-interacting DOS,
$\rho_{0}(\mu_{0})$ vanishing. This does not correspond to the physical
situation of a magnetic transition induced by the RKKY interaction. \\
{\it (ii) Difficulty in describing the criticality of transport 
properties:}  
Within the mean-field approach, the system is either in 
a strong coupling regime, with $r\neq 0$ and a heavy Fermi liquid ground
state, or in a fully decoupled regime, with $r=0$. Since
the transport properties are governed by the conduction
electrons, no universal non-Fermi-liquid behavior can be predicted at
the mean-field level. 

These two difficulties can be understood from the large$-N$
formulation of the mean-field approximation: 
Coleman~\cite{Coleman-1983}, and Read, Newns and 
Doniach~\cite{Read-Newns-Doniach-1984} have shown that magnetic
ordering involves at least processes of order $1/N$, i.e.,
fluctuations around the mean-field. 
Describing the criticality obtained from calculations 
taking into account the fluctuations and the emerging compact gauge
field theory is beyond the scope of 
this presentation. We simply refer to the works of 
Senthil, Sachdev, and Vojta~\cite{Matthias-QCP1, Matthias-QCP2}, 
who analysed the possibility of a quantum phase transition between 
a Fermi liquid phase with a large Fermi surface, and a (partially) 
fractionalized Fermi liquid phase. 
A Kondo breakdown quantum critical point has also been identified
by Pepin~\cite{Catherine-QCP, Catherine-Indranil}, who obtained, 
for example, a specific exponent diverging logarithmically 
in temperature, as observed in a number of heavy fermion metals. 

Whilst {\it Difficulty (ii)} can not be easily cured without the 
fluctuations, we describe
here how {\it Difficulty (i)} can be fixed at the mean-field level. 
The Kondo lattice Hamiltonian~(\ref{KLHamiltonian}) is 
generalized to a so-called Kondo-Heisenberg lattice, 
by adding a supplementary superexchange term, as follows: 
$H\mapsto H+\sum_{ij}J_{ij}{\bf S}_{i}{\bf S}_{j}$. 
This general model has been introduced first by 
Sengupta and Georges~\cite{Sengupta-Georges}, and studied later within
various methods. 
In terms of auxiliary fermions, introduced in 
Section~\ref{sectionmeanfield}, the superexchange term is 
rewritten as $J_{ij}{\bf S}_{i}{\bf S}_{j}=
\frac{J_{ij}}{2}\sum_{\sigma\sigma'}
f_{i\sigma}^{\dagger}f_{i\sigma'}f_{j\sigma'}^{\dagger}f_{j\sigma}$, 
which describes the spin-flip processes between two Kondo impurities,
on sites $i$ and $j$. Up to now, this description is very general and
holds for any magnetic coupling $J_{ij}$, which can be periodic (ferro
or antiferromagnetic), or randomly distributed (disorder
case). 
We will now describe two complementary mean-field approaches for the 
superexchange. 

The first one was introduced by Coqblin 
{\it et al.}~\cite{Kondo-Heisenberg-Coqblin} in the case of an 
antiferromagnetic nearest neighbor exchange $J_{ij}=J_{AF}<0$. 
In the
paramagnetic Kondo phase, the intersite exchange is approximated within 
a Resonant Valence Bound decoupling: 
$J_{ij}{\bf S}_{i}{\bf S}_{j}\mapsto 
\Gamma_{ij}\sum_{\sigma}f_{i\sigma}^{\dagger}f_{j\sigma}$, with 
$\Gamma_{ij}=\frac{J_{AF}}{2}\sum_{\sigma}
\langle f_{i\sigma}f_{j\sigma}^{\dagger}\rangle$. 
The magnetic, Heisenberg-like, interaction generates 
an effective self-consistent dispersion for the $f-$fermions. 
Using this method, Coqblin {\it et al.} show that the Kondo effect 
disappears abruptly for low band filling and/or strong intersite 
coupling~\cite{exhaustionandRKKY1, exhaustionandRKKY2}.

\begin{figure}[hhh]
\includegraphics[width=7cm, origin=br,angle=90]{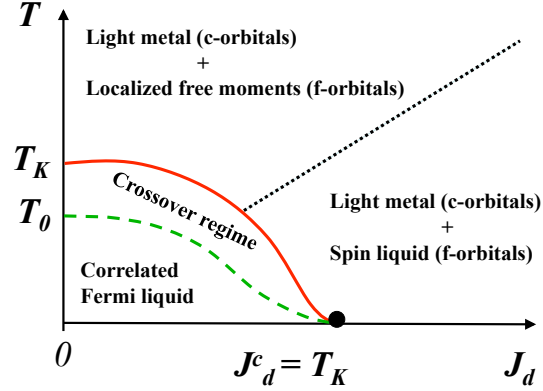}
\caption{(Color online) Schematic phase diagram of the 
disordered Kondo-Heisenberg model in the $J_{d}-T$ plane. Kondo temperature 
(red solid line) and coherence temperature (green dashed line) 
as function of $J_{d}$ are shown for fixed values of $J_{K}$ and 
$n_{c}$ in the case of $T_{0}<T_{K}$. The system is a heavy Fermi liquid 
below $T_{0}(J_{d})$. Above the line $T_{K}(J_{d})$, the 
localized spins are essentially free for $J_{d}<T$, whilst 
forming a highly correlated spin liquid for $J_{d}>T$. 
All the lines represent crossovers. 
Figure from the analysis of Ref.~\cite{BGG-2002}
}
\label{SchemasphasediagramQCP}
\end{figure}

We have developed another mean-field method~\cite{BGG-2002}, 
based on the DMFT, which we first applied to  
a disordered system where $J_{ij}$ are given 
by a Gaussian distribution with an average $[J_{ij}]=0$, and a
variance $[J_{ij}^{2}]\equiv J_{d}^{2}$.  
In this case, the superexchange term generates a local
energy-dependent self-energy for the $f-$fermions, which is determined 
by self-consistent relations similar to the ones obtained by 
Sachdev and Ye for a pure spin disordered system~\cite{Sachdev-Ye}. 
We have also applied this method to a model with constant
nearest-neighbor antiferromagnetic exchange~\cite{BGG-2003}, where 
the self-consistent equation for the $f-$fermions self-energy depends
on the lattice structure. In both cases, disordered or periodic model, 
we obtained a quantum critical point corresponding to the breakdown of 
Kondo effect, characterized by a vanishing of 
the effective hybridization, $r(T=0)=0$. 
Figure~\ref{SchemasphasediagramQCP} depicts the phase diagram that we 
obtained for the disordered Kondo-Heisenberg model~\cite{BGG-2002}. 
We also have obtained a similar phase diagram for the periodic, i.e., 
non-disordered, case~\cite{BGG-2003}. 
Here, one relevant point is that the quantum critical point emerging from 
our mean-field approach does not necessarily correspond to the onset
of magnetic ordering. This suggests an interpretation in terms of topological 
transition, without breaking of symmetry, but with a violation of 
Luttinger theorem, as discussed in 
Refs.~\cite{Matthias-QCP1, Matthias-QCP2}. 
Furthermore, $1/N$ corrections can reveal a magnetic 
instability at a value of the coupling, $J_{d}$ which is smaller than 
$J_{d}^{c}=T_{K}$. In this latter case, a symmetry breaking is
expected. 
Whereas criticality in heavy fermions is governed by a Kondo 
breakdown critical point or by a magnetic transition, what is the
nature of the non-Kondo phase, and whereas the Kondo breakdown
coincides or not with magnetic ordering are still open questions. 
The answers probably depend on the system. 

The quantum critical transition obtained here results from the 
competition between the Kondo effect and the fluctuations of the 
$f-$fermions. These fluctuations are precisely the microscopical
mechanism which is required to fix  
{\it Difficulty (i)}, i.e., 
the difficulty in driving the system to a second order
magnetic transition within the mean-field approximation. 
In the approach introduced by Coqblin {\it et al.}, the $f-$fermion 
fluctuations are generated by an effective dispersion, which is  
connected to the Resonant Valence Bound decoupling. In our approach, 
the fluctuations are included within a self-consistent local
self-energy. 
In both approaches, fluctuations are characterized by an energy,
$J_{RKKY}=J_{AF}$ or $J_{d}$, 
and the Kondo effect disappears above the critical 
value $J_{RKKY}\approx T_{K}$.

\section{Conclusions}
Important properties of Kondo systems can be obtained from 
the mean-field approximation. Some of them have not been presented
here, like, for example, the effect of a 
pseudo-gap~\cite{Withoff-Fradkin, Pseudogap-LargeN}. 
Also, for the sake of clarity, the description was restricted to 
the `standard` Kondo model. Of course, the mean-field approximation 
has been generalized and applied to more realistic models, with, 
for example, a momentum-dependent hybridization between conduction 
electrons and $f$ ions~\cite{momentumhybridization}. 

Here we have focused on the low energy scales: 
the Kondo temperature, $T_{K}$, 
characterizing the temperature crossover below which conduction
electrons and local moments are strongly coupled; the coherence 
energy, $T_{0}$, characterizing the Fermi liquid ground state; 
and $J_{RKKY}$, the intersite magnetic correlation energy. 
At the mean-field level, 
$J_{RKKY}$ is negligible or neglected, and any study of
magnetic criticality requires the inclusion of fluctuations 
around the mean-field. Nevertheless, the 
Kondo effect can break down at the mean-field level if the 
RKKY interaction is added `by hand`, generalizing the Kondo lattice to a 
Kondo-Heisenberg model. 

For the `pure` Kondo lattice, the mean-field approximation provides 
explicit expressions for $T_{0}$ and $T_{K}$. Both quantities depend
on the Kondo coupling with the same exponential factor, 
$e^{-1/J_{K}\rho_{0}(\mu_{0})}$. This explains the strong sensitivity 
of these energy scales with respect to changes in the system 
(e.g. doping, or pressure). The ratio $T_{0}/T_{K}$, which does not 
depend on $J_{K}$, appears to be a promising quantity for analyzing 
universal behaviors of heavy fermion compounds. Whilst it is equal to 
one in dilute systems, $T_{0}/T_{K}$ depends in fact on the electronic
filling, $n_{c}$, the band shape, and the impurity concentration, $x$. 
A universal regime with $T_{0}\ll T_{K}$ is expected in the 
exhaustion limit, $n_{c}\ll x$, which might be difficult to access 
experimentally. More typical situations correspond either to 
the universal dilute regime, $n_{c}\gg x$, or to the non-universal 
dense regime $x\sim n_{c}$. In the latter case, the shape of the
non-interacting DOS becomes relevant, and can lead to 
$T_{0}\ll T_{K}$ if the chemical potential, $\mu_{0}$, 
is close to a local maximum of the DOS, or to $T_{0}\gg T_{K}$ if 
$\mu_{0}$ is close to a local minimum. 

The experimental determination of $T_{0}$ and $T_{K}$ is
straightforward for systems which are `deeply` in a 
Fermi liquid phase: for example, one can determine $T_{0}$ from the 
specific heat Sommerfeld coefficient, and $T_{K}$ from the temperature 
dependence of the magnetic part of the entropy. 
The determination of $T_{K}$ might become more tricky in the vicinity 
of a magnetic transition, where the freezing of the entropy is no 
longer due to Kondo singlet formation, but to magnetic intersite
correlations instead. In this case, one should find another 
physical observable which would enable an unambiguous 
determination of $T_{K}$. In this case,  a systematic experimental 
analysis of the ratio $T_{0}/T_{K}$ might reveal interesting 
universal behavior. 

Some of the results obtained at the mean-field level have been 
confirmed by exact numerical methods. For example, DMFT calculations 
have shown that $T_{0}/T_{K}$ does not depend on the Kondo coupling. 
One limitation of the mean-field is also well known: its weakness 
in describing criticality, where fluctuations become important. 
Nevertheless, some predictions of the mean-field have not been checked
yet with more accurate methods (DMFT or $1/N$ corrections). This is,
for example, the case of the band shape effect.

\begin{acknowledgement}
I thank the organizers of the NATO
Advanced Research Workshop on Properties and Applications of 
Thermoelectric Materials. 
I acknowledge my collaborators on the works presented here:   
P. Fulde, A. Georges, M. Grilli, and V. Zlatic. 
I am also grateful to A. Rosch, M. Vojta, A. Klopper, H. Weber and N.B. Perkins for 
useful discussions regarding this manuscript, and to P. Nozi\`eres, 
B. Coqblin, and C. Lacroix for fruitful advises.

\end{acknowledgement}

%
%
\end{document}